\documentstyle[psfig]{mn}
\begin{document}
\input epsf
\newcommand{\msun}{M$_{\odot}$}
\newcommand{\lsun}{ML$_{\odot}$}
\newcommand{\halpha}{${\rm H\alpha}$}
\newcommand{\bd}{\mbox{BD -${\rm 22^{\circ} 3467}$}}
\newcommand{\bbd}{\mbox{${\rm \bf BD -22^{\circ} 3467}$}}
\newcommand{\abell}{Abell 35}
\newcommand{\jnw}{Jasniewicz}
\newcommand{\arc}{$^{\prime\prime}$}
\newcommand{\arcm}{$^{\prime}$}
\newcommand{\km}{km s$^{-1}$}
\newcommand{\dg}{^{\circ}}
\def\gape{\mathrel{\raisebox{-.9ex}{\mbox{$\stackrel{\textstyle >}{\sim}$}}}}
\def\lape{\mathrel{\raisebox{-.9ex}{\mbox{$\stackrel{\textstyle <}{\sim}$}}}}
\title
{The separation of the stars in the binary nucleus of the planetary nebula 
Abell 35} 
\author[A. A. Gatti et al]
{ A. A. Gatti$^{1}$, J. E. Drew$^{1}$,
R. D. Oudmaijer$^{1}$, T. R. Marsh$^{2}$, A. E. Lynas-Gray$^{3}$\\
$^{1}$ Imperial College of Science, Technology and Medicine,
Blackett Laboratory, Prince Consort Road, London,  SW7 2BZ, U.K.   \\
$^{2}$ Department of Physics, Southampton University, Southampton S09
5NH, U.K. \\
$^{3}$ Department of Physics, University of Oxford, Nuclear Physics
Laboratory, Keble Road, Oxford, OX1 3RH, U.K.} 
\date{received,  accepted}
\maketitle
\begin{abstract}
Using the Planetary Camera on board the Hubble Space Telescope we have 
measured the projected separation
of the binary components in the nucleus of the planetary nebula Abell 35 to be
larger than 0.08$^{\prime\prime}$ but less than 0.14$^{\prime\prime}$.
The system was imaged in three filters centered at 
2950${\rm \AA}$, 3350${\rm \AA}$ and 5785${\rm \AA}$. The white dwarf primary 
star responsible for ionizing the nebula is half as bright as its companion
in the 2950${\rm \AA}$ filter causing the source to be visibly
elongated. The 3350${\rm \AA}$ setting, on the other hand, shows no
elongation as a result of the more extreme flux ratio.
The F300W data allows the determinination of the binary's
projected separation.  At the minimum distance of 160 parsec to the system, 
our result corresponds to 18$\pm 5$ AU. This outcome is consistent
with the wind accretion induced rapid rotation hypothesis, but cannot be
reconciled with the binary having emerged from a common-envelope phase.
\end{abstract}
\begin{keywords}
ISM: planetary nebulae: individual (Abell 35)
binaries: close
binaries: general
stars: evolution
stars: peculiar
\end{keywords}

\section{Introduction}
The planetary nebula Abell 35
($\alpha_{50}$=12$^{h}$50.9$^{m}$, $\delta_{50}$=$-22^{o}$36$^{\prime}$;
l=303$^{o}$, b=$+40^{o}$) was hypothesized to
possess a binary nucleus by Jacoby (1981) to account for the apparent
lack of an ionizing source and for the blue excess of LW Hya (\bd), 
the bright G8 III-IV
star, visible off-center within the nebula. The white dwarf primary 
was first reported in IUE SWP spectra taken by Grewing and Bianchi (1988). 

\vspace{0.2cm}

\begin{figure*}
\centerline{\psfig{figure=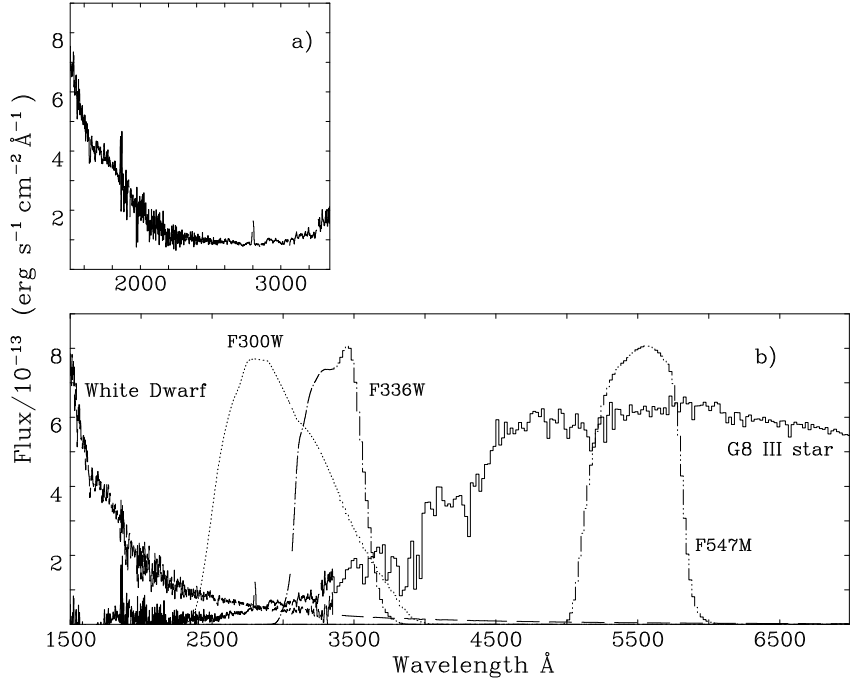,angle=-90,width=12.cm}}
\caption[]{a) the SWP and LWP IUE spectrum of the central star of Abell
35. b) The IUE spectrum decomposed into its white dwarf and G star
contributions (see text for details). Also shown are the relevant WFPC2
filter profile functions. The relative flux contributions
of the two components in the different filters were calculated by 
folding the filter profiles with the single-star spectra and comparing.} 

\end{figure*} 

\begin{figure*}
\centerline{\psfig{figure=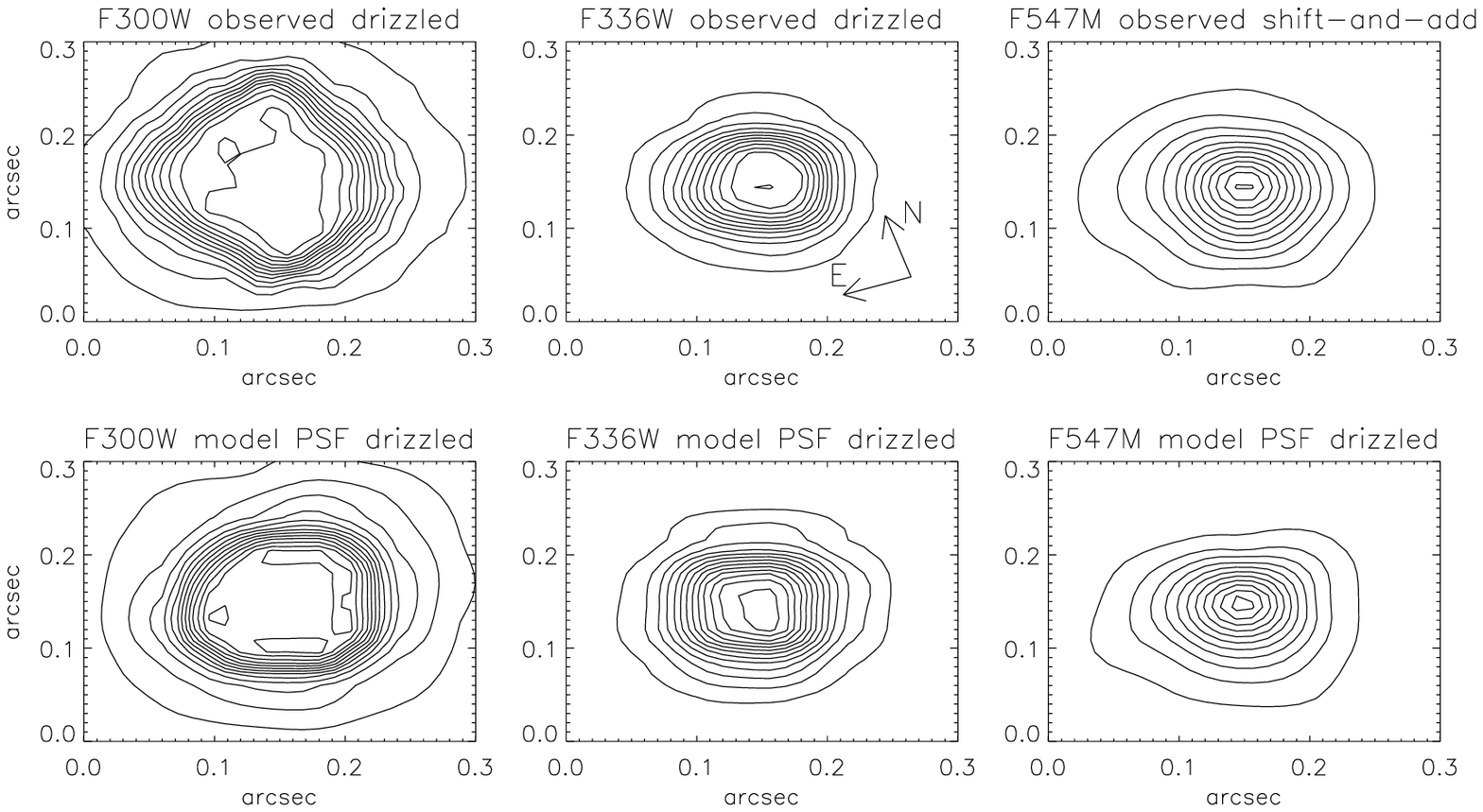,width=15.cm,height=11.cm}}
\caption[]{The observed images (top) and the TINYTIM produced 
model point spread functions (bottom) for all three filters: 
F300W, F336W and F547M. The F300W and F336W have
been ``drizzled'' to a pixel size of 22.75 mas, half the original PC 
size. The F547M images, on the other hand, retain the original PC size.
The F300W model image has been saturated to the known level. 
The observed F300W image is clearly elongated 
with respect to the PSF along a $\sim$NNE--SSW axis that is almost 
perpendicular to a slight elongation apparent in the PSF itself.  
The contour levels indicate $6\%$ flux steps.}
\end{figure*}

Because of a number of extraordinary features, Abell 35 has
represented a  challenge to theories of binary star
evolution. More specifically:
\begin{enumerate}
\item
The optically-dominant cool component, LW Hya, is
distinguished from normal G stars by its remarkably high projected
rotational velocity (vsin~i=90 km s$^{-1}$, Vilhu, Gustaffson \& Walter, 
1991). G type
stars with such large vsin~i are rare and usually classified as FK
Comae objects. The origin of FK Comae stars is unclear, although it
has been suggested that they are either coalesced/coalescing contact
binaries (e.g. see Welty \& Ramsey, 1994) or the remnants
of a mass transfer event with a thus far undetected white dwarf companion
(Walter \& Basri, 1982). 
\\
\item
Long-term radial velocity measurements of \bd\ have shown no evidence
of variations due to
orbital motion (Gatti et al., 1997), suggesting that an orbital period in
excess of 10 years could well turn out to be appropriate.   
\\
\item
Based on their low-dispersion UV spectrum, Grewing and Bianchi (1988)
noted that the white dwarf central star bears close resemblance to the
class of PG 1159 objects (Werner et al. 1996). These are rapidly 
pulsating hot stars which have just joined or are about to join the 
white dwarf cooling track. 
\end{enumerate}
LoTr 1 and LoTr 5 (Longmore \& Tritton, 1980) are the only
other planetary nebulae believed to
have similar central star binaries to that in Abell 35. The three systems are
therefore often referred to as the Abell 35-like objects.
\begin{table}
\begin{center}
\caption[]{Log of exposures. Sub-pixel offsets were applied for
``drizzling'' purposes to both the F300W and the F336W filters.}
\vspace{0.12cm}
{\small
\begin{tabular} {|c|c|c|}\hline
Filter&Exp. time&HJD \\ \hline
F300W&60.0s&2450824.107 \\ 
F300W&60.0s&2450824.110 \\ 
F300W&60.0s&2450824.112 \\ 
F300W&60.0s&2450824.115 \\ \hline
F336W&12.0s&2450824.121 \\ 
F336W&12.0s&2450824.123 \\ 
F336W&12.0s&2450824.125 \\ 
F336W&12.0s&2450824.127 \\ \hline
F547M&0.4s&2450824.131 \\ 
F547M&0.4s&2450824.133 \\ 
F547M&0.4s&2450824.135 \\ 
F547M&1.0s&2450824.136 \\ 
F547M&1.0s&2450824.137 \\ \hline
\end{tabular}}
\end{center}
\end{table}

\vspace{0.20cm}

Two evolutionary scenarios have been invoked to explain the Abell-35 like
nuclei. The first of these is the ``common envelope'' hypothesis (Bond
and Livio, 1990) according to which the binary's primary has undergone 
dynamically unstable Roche Lobe overflow. The second, the ``wind accretion
induced rapid rotation'' model where slow
accretion of the AGB wind from the white dwarf progenitor in a
detached configuration accounts for the angular momentum transfer
that spins up the companion (eg. Jeffries \& Stevens, 1996 and
references therein). Since the common envelope scenario involves the
primary star expanding to fill its Roche Lobe, while the wind accretion
model applies to a fully-detached configuration, a distinction between 
these models may arise from a clear determination of the 
present-day binary separation. 

As part of a wider programme of observation of the nucleus of Abell 35,
we have obtained HST Planetary Camera (PC) images of it spanning the UV and
optical domain, with a view to finding out if the binary is spatially
resolvable. If it is, we can establish a lower limit on the radius of
the binary orbit which may rule out the common envelope hypothesis. The
pixel size of the PC is 45.5 mas which
corresponds to a projected separation of 8 AU at 
160 pc, the likely minimum distance to Abell 35 (see section 4). 
Prior to this work, the separation was only constrained to be below
$\sim$ 1 arcsec from IUE data. We will now present HST PC data
which shows the binary to be resolvable.

\section{Observations}

In January 1998, the Hubble Space Telescope Wide Field PC 2 
was used to obtain 13 exposures (Table 1) of the binary nucleus of
Abell 35 in three filter settings: F300W (${\rm \lambda_{c} \sim 2950
\AA}$, $\lambda\lambda \sim$ 2300--4000 \AA ), F336W 
(${\rm \lambda_{c} \sim
3350 \AA}$, $\lambda\lambda \sim$ 2900--3800 \AA ) and F547M (${\rm
\lambda_{c} \sim 5785 \AA}$, $\lambda\lambda \sim$ 5000 -- 6000 \AA ). 
Based on the IUE SWP and LWP spectra of
the system (Grewing and Bianchi, 1988; Jasniewicz et al. 1994), 
the G8 III-IV star starts to dominate the
total flux longwards of 2800~\AA . \\
To determine quantitatively the relative
flux of the two stellar components in each filter, the IUE
spectra were decomposed for its white dwarf and G-star contributions
by comparison with a PG1159-type white dwarf IUE spectrum (fig 1). The 
result was then extrapolated to higher wavelengths assuming a Rayleigh-Jeans
wavelength dependence for the white dwarf and using either a 5000K or a
5500K, log g=3.5, Kurucz model atmosphere normalized to V=9.6mag for 
the G-star. The resulting single-star spectra for both
components were then folded with the WFPC2 filter profile
functions. Depending on whether a temperature of 5000 or 5500K is 
assumed for the G star, we find the expected G star 
to white dwarf flux ratio to be: 1.8--2.2 for the F300W filter, 
5.3--6.2 for the F336W filter and 115--122 for the F547M filter. 

\vspace{0.1cm}

Both the F300W and F336W exposures were 
sub-stepped by a non-integral pixel amount to allow the recovery of
high frequency spatial information lost by the factor of two
undersampling of the PSF by the pixels of the PC. These multiple offset
images were linearly reconstructed using the DRIZZLE software written 
by Fruchter \& Hook (1996). The validity of the reconstructed, higher 
resolution, images was assessed in both filters by comparison with a
reconstruction performed without change in pixel size (i.e. by using
the DRIZZLE routine to simulate a straightforward shift-and-add
procedure). The F547M exposures were simply co-added after 
correcting for cosmic rays. 
In all cases the root-mean-square jitter of the telescope was found to be 
$\le$ 7.0~mas, indicating no large excursions of the tracking during
the individual exposures.

Further, for each filter, model point spread functions to be compared
with the data were generated with the TINYTIM software (Krist, 1995). A
number of subsampling parameters were used and, when necessary, 
the resulting models were rescaled and convolved with the pixel
response kernel.
\begin{figure*}
\centerline{\psfig{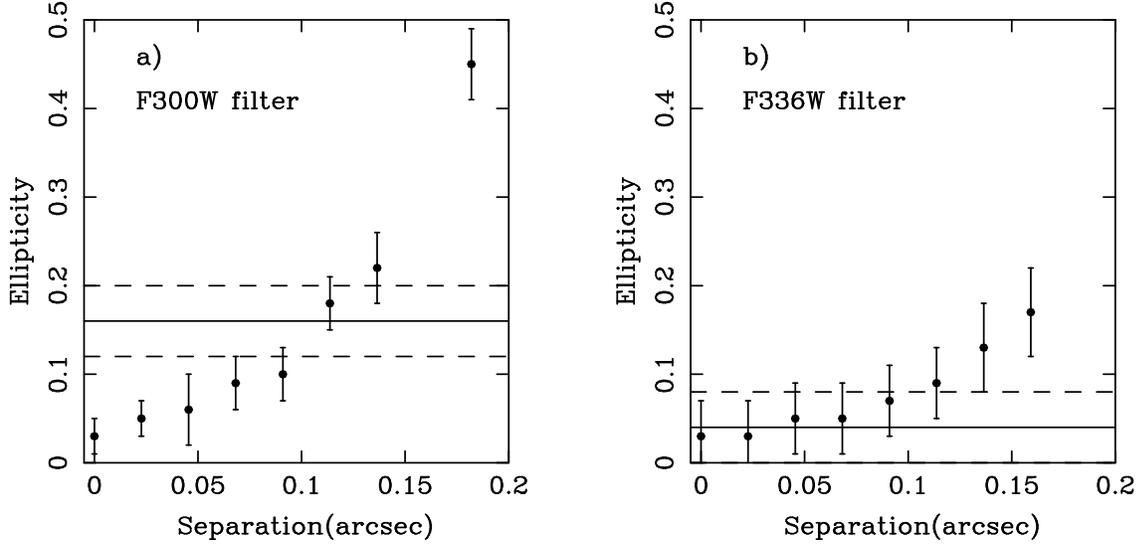}}
\caption[]{a) The ellipticity as a function of separation for F300W
model images produced with TINYTIM. Each value is the 
average ellipticity for 8 models with fixed separation and varying 
position angles encompassing all 360$\dg$ in 45$\dg$ steps. 
The 3$\sigma$ error bars reflect the uncertainty in the
measured ellipticity as a function of position angle mainly due to
pixellation effects. The solid line is the measured ellipticity
of the observed F300W image with the associated error. Only separations in the
range 0.08-0.14 arcsec are consistent with the observations. Smaller
separations cannot account for the elongation of the F300W
data. The non zero ellipticity of a single source point spread
function reflects its slight elongation in a direction almost
perpendicular to the observation (fig 2). Panel b) shows the result of
the analysis applied to the F336W images. In this circumstance, the
observations are consistent with a binary source with orbital 
separation $\leq 0.14^{\prime\prime}$, consistent with the upper limit
produced by the F300W data.} 
\end{figure*} 

\section{Results}

The leftmost two panels of figure 2 show the F300W drizzled image and the
PSF as modelled by TINYTIM for a G8 spectral type object (which does not
differ noticeably from the PSF appropriate to a hot white dwarf).  
It can be seen immediately that there is an
elongation of the observations in an approximately NS direction which
cannot be directly related to an artifact of the PSF. 
However, because the  F300W image is saturated (due to A/D conversion
only, and not to blooming), it is necessary to assess whether the
elongation could have been induced by an optical error in the F300W 
filter made more acute by effect of the saturation. We have done this 
by producing artificial images for a number of source 
separations and position angles. To compare with the data 
we have scaled the relative counts of the two components
to the flux ratio calculated from the IUE spectra and saturated the
images to the known level. We have then drizzled the images following 
the same procedure as with the observations. 

A meaningful comparison of the simulated data with the observations is 
not trivial. Mainly as a result
of pixellation effects and because of the intrinsic asymmetry of the
WFPC2 PSFs, the overall shape of the calculated image and its resulting center 
change subtly as a function of the position angle of the two
sources. A $\chi^2$ minimization procedure cannot be used as a
similarity test because of its strong dependence on an accurate
alignment between models and observations. The overall shape
of the model images, on the other hand, can. We have quantified this by 
measuring the image ellipticity as a function of separation for the models 
and comparing this with the F300W observations (fig 3a). Ellipticity 
measurements are independent of the image centers but vary slightly 
with assumed position angle (again mostly due to pixellation
effects). Models spanning position angles 0-315 in steps of 45 degrees 
were made and the average ellipticity was calculated. The error bars
shown in fig 3 are 3$\sigma$, and were determined from the ellipse
fitting routine. They also reflect the scatter with changing position
angle. The ellipticity of the F300W data was found to be 0.16$\pm$0.04, 
inconsistent with the data arising from a single, unresolved, point
source with a plausible PSF. In fact, only separations in 
the range 0.08-0.14 arcseconds can be reconciled with the observed
F300W image. In view of the 3$\sigma$ significance of the error bars 
used here, this must be regarded as a maximum allowable range.

\vspace{0.2cm}

This result was further checked against the F336W
data. The lack of any detectable elongation in this filter
(fig 2, middle panels) is not surprising in view of the more extreme flux 
ratio and modest separation of the sources, suggested by the F300W
data. We have modelled the F336W data 
following the same procedure as for the F300W observations, varying the
separation and adopting a flux ratio of 5 (which maximizes the white 
dwarf flux contribution). This ellipticity test
reveals the F336W filter to be far less sensitive to
the binary separation. In fact, we find that no significant elongation 
appears in the overall F336W image until a separation of the order of 
0.12--0.14 arcsecs is reached (fig 3b). The F336W observation therefore sets
the same upper limit of 0.14 arcseconds on the separation of the
binary components as the F300W data.

\vspace{0.2cm}

Hence, on the basis of the F300W data, we conclude that the projected
separation of the binary components falls in the range 0.08--0.14
arcseconds.

\section{Discussion and Conclusion}
\begin{table}
{\small
\begin{center}
\caption[]{Estimates of the distance to Abell 35}
\begin{tabular} {|c c c|}
\hline\hline
D(pc) (ref)&method&comments \\ \hline
240 (1)&${\rm H_{\alpha } + [NII]}$ line flux &---   \\ \hline
208 (2)&5 GHz radio flux&--- \\ \hline
$360 \pm 80$ (3)&photometry&--- \\ \hline
$\sim 310$ (4)& blackbody fit to WD & $M_{WD}=0.6 M_{\odot}$  \\ \hline
$134^{+33}_{-23}$ (5)&Trigonom. Parallax & uncorrected for bias \\ \hline
\end{tabular}
\end{center}}
$^{1}$Abell, 1966;$^{2}$ Milne, 1979;$^{3}$Jacoby, 1981;$^{4}$Hollis et al.,
1996; $^{5}$ ESA, 1997
\end{table}

The translation of the derived separation of
into a projected orbital separation
requires a reliable distance to Abell 35. This is not yet available.
In table 2 we present a short summary of the distance
estimates to Abell 35 currently in the literature. 
 
The recent Hipparcos measurement ($\pi$ = 7.48 $\pm$ 1.55 mas) seems to
suggest a substantial decrease in the distance compared to previous
estimates.  However, the value in table 2 should be corrected for the
Lutz-Kelker statistical bias which leads to an under-estimate of
the distance to an object (Lutz \& Kelker 1973; Koen 1992; 
see also Oudmaijer, Groenewegen \& Schrijver 1998).
Koen (1992) calculated the bias analytically and provided 90$\%$
confidence intervals of the bias.  Although the relative error of
$\sim 20\%$ in the observed parallax for LW Hya is just too large to be
calculated analytically, we may gain some insight by applying the bias
correction for a 17.5$\%$ error (Koen's maximum value) to the data of
Abell 35.  This leads to a (mean) bias corrected distance of
163$^{+96}_{-58}$ pc. In view of the earlier distance estimates, we
adopt 160 pc as a {\it minimum} distance. 

\vspace{0.2cm}

At 160 pc, the measured separation between the white
dwarf and the G-star translates into a {\it projected} orbital
separation of 18$\pm$5 AU. For this minimum value of the orbital
separation, and a likely maximum total mass of the system 
of $\sim$ 3\msun\, we find that the orbital period for a circular 
orbit about the system's barycenter is P $>$ 40 years. This 
estimate is consistent with Gatti et al.'s (1997)
conclusion that the binary period must be longer than 10 years. 
Should the orbit turn out to be significantly eccentric, our minimum 
period estimate would not apply.

\vspace{0.2cm}

The results presented in this paper exclude the possibility that Abell 35 
may be the remnant of a common-envelope phase.  For example, if we assume a 
relatively favourable initial configuration of a 5\msun\ AGB primary and a 
1\msun\ secondary at the minimum acceptable orbital separation of 13~AU 
(obtained for the minimum distance of 160~pc and minimum projected
separation of 0.08 arcsec), we can estimate the Roche Lobe radius of the 
primary to be $\sim$7~AU (Eggleton, 1983).  This figure is itself a minimum
given that the common-envelope phase is expected to result in orbit shrinkage.
Since no AGB star can reach a 7~AU radius without breaking the core 
mass-luminosity relation (Paczynski, 1970), the primary star in Abell 35 
cannot have filled its Roche Lobe. 

On the other hand, by inspection of the model results presented by 
Jeffries \& Stevens (1996) we find that, at separations matching or
exceeding the minimum deduced from our data, the spin up of the companion 
star \bd\ to a period of 18 hours (Jasniewicz \& Acker, 1988) is easily 
achieved by accreting a few percent of a solar mass. This is entirely 
consistent with values for wind accretion rates obtained in hydrodynamical 
models of binary stars (Theuns, Boffin \& Jorissen 1996 and references 
therein). Important tasks for the future are to test wind-accretion models
more precisely by, for example, obtaining constraints on the present-day 
component masses in this intriguing binary.

\section*{Acknowledgments}
The authors are extremely grateful to Jeremy Walsh and Richard Hook 
for their help in the analysis of the WFPC2 images. We would also like
to thank the referee, George Jacoby, for his useful comments. AAG acknowledges 
financial support from the Knight Award of the University of London, from the 
Department of Physics of Imperial College and from the University of
Rome `La Sapienza'.  This work is based on observations with the NASA/ESA
Hubble Space Telescope obtained at the Space Telescope Science Institute,
which is operated by AURA Inc. under contract to NASA.


\begin{thebibliography}{99}
\bibitem{b1} Abell G. O., 1966, ApJ, 144, 259
\bibitem{b2} Bond H. E., Livio M., 1990, ApJ, 355, 568 
\bibitem{b3} Eggleton P. P., 1983, ApJ, 268,368 
\bibitem{b4} ESA, 1997, The Hipparcos and Tycho Catalogues, ESA SP-1200
\bibitem{b5} Fruchter A., R. Hook, 1996, Applications of Digital Image
Processing XX, ed. A. Tescher, Proc. S.P.I.E. vol 3164 
\bibitem{b6} Gatti A. A., Drew J. E., Lumsden S., Marsh T., Moran C.,
Stetson P., 1997, MNRAS, 291, 773
\bibitem{b67} Grewing M., Bianchi L., 1988, in {\sl A Decade of UV Astronomy 
with IUE}, ESA SP-281, 2, 177
\bibitem{b8} Hollis J. M., Van Buren D., Vogel S. N., Feibelman W. A.,
Jacoby G. H., Pedelty J. A., 1996, ApJ, 456, 644
\bibitem{b9} Jacoby G. H., 1981, ApJ, 244, 903
\bibitem{b10} Jasniewicz G., Acker A., 1988, A\&A, 189, L7
\bibitem{b11} Jasniewicz G., Lapierre G., Monier R. 1994, A\&A, 287, 591
\bibitem{b12} Jeffries R. D., Stevens I. R., 1996, MNRAS, 279, 180
\bibitem{b13} Koen, C., 1992, MNRAS, 256, 65
\bibitem{b14} Krist J., 1995, in {\sl Astronomical Data Analysis Software and
Systems IV}, eds. Shaw R. A., Payne H. E., Hayes J. J. E., 
ASP Conf.Ser.No. 77, 349
\bibitem{b15} Longmore A. J., Tritton S. B., 1980, MNRAS, 193, 521
\bibitem{b16} Lutz T. E., Kelker D. H., 1973, PASP, 85, 573 
\bibitem{b17} Milne D. K., 1979, A\&AS, 36, 227
\bibitem{b18} Oudmaijer R. D., Groenewegen M. A. T., Schrijver H., 1998, 
MNRAS, 294, L41 
\bibitem{b19} Paczynski B., 1970, Acta Astron. 20, 47
\bibitem{b20} Theuns T., Boffin H. M. J., Jorissen A., MNRAS, 280, 1264
\bibitem{b21} Vilhu O., Gustafsson B., Walter F. M., 1991, A\&A, 241,167
\bibitem{b22} Walter F. M., Basri G. S., 1982, ApJ, 260, 735
\bibitem{b23} Welty A. D., Ramsey L. W., 1994, ApJ, 435, 848
\bibitem{b24} Werner K., Dreizler S., Heber U., Rauch T., 1996, in
{\sl Hydrogen Deficient Stars}, eds. Jeffery C. S.,
Heber U., ASP Conf.Ser.No. 96, 267
\end{thebibliography}
\end{document}